# Epitaxial Ferroelectric La-doped $Hf_{0.5}Zr_{0.5}O_2$ Thin Films

Tingfeng Song,[1] Romain Bachelet,[2] Guillaume Saint-Girons,[2] Raul Solanas,[1] Ignasi Fina,[1,*] Florencio Sánchez[1,*]

[1] Institut de Ciència de Materials de Barcelona (ICMAB-CSIC), Campus UAB, Bellaterra 08193, Barcelona, Spain
[2] Institut des Nanotechnologies de Lyon (INL-CNRS UMR 5270), Université de Lyon, Ecole Centrale de Lyon, 36 avenue Guy de Collongue, 69134 Ecully Cedex, France.

ABSTRACT

Doping ferroelectric $Hf_{0.5}Zr_{0.5}O_2$ with La is a promising route to improve endurance. However, the beneficial effect of La on the endurance of polycrystalline films may be accompanied by degradation of the retention. We have investigated the endurance - retention dilemma in La-doped epitaxial films. Compared to undoped epitaxial films, large values of polarization are obtained in a wider thickness range, whereas the coercive fields are similar, and the leakage current is substantially reduced. Compared to polycrystalline La-doped films, epitaxial La-doped films show more fatigue but there is not significant wake-up effect and endurance-retention dilemma. The persistent wake-up effect common to polycrystalline La-doped $Hf_{0.5}Zr_{0.5}O_2$ films, is limited to a few cycles in epitaxial films. Despite fatigue, endurance in epitaxial La-doped films is more than $10^{10}$ cycles, and this good property is accompanied by excellent retention of more than 10 years. These results demonstrate that wake-up effect and endurance-retention dilemma are not intrinsic in La-doped $Hf_{0.5}Zr_{0.5}O_2$.

INTRODUCTION

Stabilization of the metastable orthorhombic (o) ferroelectric phase in doped $HfO_2$ films[1] can have a great technological impact. $HfO_2$ films can be grown by atomic layer deposition, a process compatible with CMOS technology, and can overcome the lack of scalability of conventional ferroelectric oxides.[2-4] Ferroelectric hafnia holds promise for non-volatile random access devices, but also for emerging ferroelectric memories such as tunnel junctions and field effect transistors. However, the endurance of $HfO_2$ capacitors, which is a critical property for memory devices, is moderately low. The low endurance of ferroelectric $HfO_2$ is perhaps the consequence of a large coercive field ($E_c$), >1 MV/cm, around one order of magnitude greater than that of perovskite oxides. The electric field required to cycle $HfO_2$ is very high, close to the breakdown field, and endurance of $HfO_2$ capacitors is generally limited by hard breakdown that usually occurs well below $10^{10}$ cycles.

Doping with trivalent La atoms has allowed a significant improvement. In particular, $Hf_{0.5}Zr_{0.5}O_2$ (HZO) films doped with 1 mol% La show increased polarization and endurance up to $4 \times 10^{10}$ cycles.[5] This was accomplished by reducing $E_c$ and leakage current by around 30% and three orders of magnitude, respectively.[5] Optimizing the amount of La has allowed a further enhancement.[6] HZO capacitors doped with 0.7 mol% La remained operational after $10^{11}$ cycles, and without polarization reduction (fatigue).[6] However, optimization of endurance by La-doping can be accompanied by a severe degradation of polarization retention.[7] The endurance of films doped with 2.5 mol% La, the optimal amount considering all ferroelectric properties, was only $10^7$ cycles. On the other hand, a common detrimental effect of La doping is the highly increased wake-up effect up to $10^5 - 10^7$ cycles.[5-7]

Studies on the effects of doping with La have been carried out with polycrystalline films. Ferroelectricity has been reported for epitaxial films of $HfO_2$ doped with several atoms, including Y,[8-9] Zr,[10-24] and Si.[25] The remanent polarization ($P_r$) in epitaxial films can exceed 20 $\mu C/cm^2$, and the $E_c$ is greater than in polycrystalline samples.[10-17, 22-25] In addition, the usual $E_c$ - thickness (t) scaling in conventional ferroelectric perovskites, $E_c \propto t^{-2/3}$,[26-27] elusive in polycrystalline HZO or other doped-$HfO_2$ films, is observed in epitaxial HZO films.[15, 22] Endurance of epitaxial HZO films can be high in spite of their enormous $E_c$ of 3-4 MV/cm.[10, 14, 16, 22] Indeed, endurance of $10^{11}$ cycles has been measured in sub-5 nm HZO films applying a very large electric field above 5 MV/cm.[22] Further improvement could be achieved by decreasing leakage and the huge $E_c$ of epitaxial films. By doping with La, an acceptor dopant for HZO, it has been possible to reduce leakage and $E_c$ in polycrystalline HZO films and, therefore, it is of great interest to stabilize the o-phase in epitaxial La-doped HZO (La:HZO) films, and measure ferroelectric polarization, endurance and retention. In order to investigate the effects of La-doping in epitaxial films, we fixed the La content to the same amount (1 mol% La) that in the pioneering study by Chernikova et al.,[5] and La-doped HZO epitaxial films were deposited using growth conditions, substrates and bottom electrode suitable to stabilize the o-phase in undoped HZO films.

Epitaxial La:HZO films of thickness (t) in the 4.5 - 13 nm range were deposited on (001)-oriented $SrTiO_3$ (STO) substrates buffered with a $La_{2/3}Sr_{1/3}MnO_3$ (LSMO) electrode. The o-phase is epitaxially stabilized, and the films have $P_r$ up to ~ 28 $\mu C/cm^2$, and $E_c$ that scales with thickness according to the $\propto t^{-2/3}$ dependence. Compared to undoped HZO epitaxial films of same thickness, $E_c$ values are similar, while leakage current reduces by around one order of magnitude. Endurance improves over equivalent undoped epitaxial films, with strong thickness dependence in both doped and undoped films. Moreover, wake-up effect (100 cycles), hardly observed in epitaxial undoped HZO films, is evident in films thinner than around 8 nm although being it much less than for polycrystalline La:HZO. Films t < 10 nm show $2P_r$ > 2 $\mu C/cm^2$ after $10^{10}$ cycles at field 4-5 MV/cm. Remarkably, high endurance is accompanied by long retention of more than 10 years using the same poling field. This demonstrates that high endurance and long retention can be attained simultaneously in La:HZO films on LSMO/STO(001). However, since Si wafers are required for applications, epitaxial La:HZO films were also grown on LSMO/STO/Si(001). LSMO and STO were grown by pulsed laser deposition (PLD) and molecular beam epitaxy (MBE), respectively. We note that STO and other perovskites can also be integrated epitaxially on Si(001) by atomic layer deposition, allowing for conformal deposition.[28] The epitaxial HZO films on Si(001) also exhibit high polarization, endurance and retention.

EXPERIMENTAL

La-doped and undoped HZO films were grown on LSMO/STO(001) and LSMO/STO/Si(001). The STO(001) single crystalline substrates were used in as-received conditions, thus presenting a majority of $TiO_2$ chemical terminations and minority of SrO.[29] The STO buffer layers, 26 nm thick, were grown *ex-situ* by molecular beam epitaxy (MBE). Additional information about MBE deposition of STO and its structural properties is reported elsewhere.[16] HZO (doped and undoped) and LSMO films were grown in a single process by pulsed laser deposition (PLD) using a KrF excimer laser. The laser frequency was 5 and 2 Hz for LSMO and HZO, respectively. The composition of the PLD targets was $Hf_{0.5}Zr_{0.49}La_{0.01}O_2$, $Hf_{0.5}Zr_{0.5}O_2$ and $La_{2/3}Sr_{1/3}MnO_3$. Deposition was done under dynamic oxygen pressure of 0.1 mbar at 700 °C and 800 °C (heater block temperature) for LSMO and HZO, respectively. LSMO layer is used as bottom electrode with t = 25 nm thick, and HZO films are in the t = 4.5 - 13 nm range, being HZO thickness determined by the number of laser pulses used in the PLD process. At the end of the deposition, samples were cooled under 0.2 mbar oxygen pressure. Capacitors were fabricated chosen platinum, which can be deposited at room temperature, as top electrode. Pt circular top electrodes, of diameter 20 μm and thickness 20 nm, deposited by sputtering through stencil masks.

Crystal structure was studied by X-ray diffraction (XRD) with Cu Kα radiation using a Siemens D5000 diffractometer equipped with point detector, and a Bruker D8-Advance diffractometer equipped with 2D detector. Surface topography was studied using atomic force microscopy (AFM) using a Keysight 5100. Ferroelectric polarization loops were measured using an AixACCT TFAnalyser2000 platform at 1 kHz by the dynamic leakage current compensation (DLCC) procedure[30-31] at room temperature in top-bottom configuration. By DLCC method, the leakage current effects is suppressed based on two assumptions: that the leakage current is frequency independent and that the dielectric current (including ferroelectric switching current) is linearly dependent on the frequency. Therefore, leakage contribution to measured current can be subtracted by measuring cycles at two different frequencies (v and v/2), and subtracting frequency independent current contribution. Endurance was measured cycling the sample at frequency of 100 kHz using bipolar square pulses of indicated amplitude and measuring polarization loops at 1 kHz. Retention was measured poling the sample using triangular pulse of 0.25 ms and determining the $P_r$ from the first polarization curve of the polarization loop measured at 1kHz using the PUND protocol after a delay time.[31]

RESULTS

Figure 1a shows the XRD θ-2θ scans of the La:HZO/LSMO/STO(001) samples. For clarity, the scans are vertically shifted, increasing HZO thickness from bottom to top. Besides the (001) and (002) reflections of the STO substrate and LSMO electrode, there is a peak at 2θ around 30°, coincident with the position of the o-HZO(111) reflection in epitaxial undoped HZO films.[15] A zoomed 2θ range around this peak, scanned with longer acquisition time, is in the right panel of Figure 1a. Increasing thickness, the o-HZO(111) peak becomes narrower and Laue oscillations are evident. The thickness of the thickest film, calculated by simulation of the Laue oscillations (Supporting information S1) is 13.1 nm. The simulation shows an asymmetry, with the satellites being more intense at lower angles than at higher angles with respect to the o-HZO(111) peak. Asymmetry is observed in similar undoped epitaxial HZO(111) films.[13,15] Laue oscillations around LSMO reflections are not observed due to the LSMO thickness (25 nm) and the measurement with Cu Kα radiation that included Cu Kα$_1$ and Cu Kα$_2$ components. Pole figures (Supporting information S2) confirm epitaxy, with four in-plane crystal variants and same epitaxial relationship as the films without La.[10, 15, 17, 21] Very low intensity peaks barely distinguished at 2θ=34° in Figure 1a, are likely m-HZO{200} reflections. XRD 2θ-χ frames (Supporting information S3) show intense o-HZO(111) spots, while the monoclinic reflections are not detected. This signals very low amount of monoclinic phase in epitaxial La:HZO films, smaller than the observed in equivalent epitaxial undoped HZO films.[15, 17, 21] Similar reduction of monoclinic phase was observed in polycrystalline La-doped HZO films.[5, 7] In summary, XRD confirms that the orthorhombic phase present in the La-doped HZO films has grown epitaxially, but exhibits crystal variants and coexists with a minority monoclinic phase. Therefore, the films are not monocrystalline: the majority orthorhombic phase presents crystal variants and coexists with the minority monoclinic phase, existing grain boundaries between crystal variants and between the two phases.

XRD θ-2θ scans of the La:HZO/LSMO/STO/Si(001) samples (Figure 1b) confirm that the orthorhombic phase is also stabilized on Si(001). There are Laue oscillations around the o-HZO(111) peak (see right panel in Figure 1b and Supporting Information S1). Traces of m-HZO{200} reflections are more evident in these films than on STO(001), and the 2θ-χ frames (Supporting information S3) recorded for the thicker films show elongated low intensity m-HZO{200} spots. Next, we determined the out-of-plane (oop) lattice parameter of the orthorhombic phase, $d_{o-HZO(111)}$, from the position of the o-HZO(111) peak in the θ-2θ scans (Figures 1a and 1b). The dependence of $d_{o-HZO(111)}$ on thickness is represented in Figure 1c, for films on STO(001) (red triangles) and Si(001) (black squares). In both series, the oop parameter decreases slightly with t up to t ~ 8 nm, and the value remains constant in the thicker film (t ~ 13 nm). Similar dependence (blue squares) was found for undoped HZO films on STO(001).[15] The smaller $d_{o-HZO(111)}$ values of the films on Si(001) respect the equivalent films on STO(001) is a consequence of the large mismatch in thermal expansion coefficients between the oxides and silicon. Due to the much smaller thermal expansion coefficient of Si, HZO will be under in-plane tensile stress when cooled after deposition, causing contraction of the out-of-plane $d_{o-HZO(111)}$ parameter. The oop parameters of the La-doped and undoped HZO films on Si(001) are quite similar, with slightly higher values in the series of doped films. The higher $d_{o-HZO(111)}$ in La doped films differs from the reported slight shrinkage for polycrystalline HZO films doped with similar La content.[5, 7]

The surface of the films is very flat. The atomic force microscopy (AFM) topographic images of La:HZO films on STO(001) (Figure 2) show terraces and steps morphology. This morphology is more evident in the t = 8.3 nm, which presents the narrower terraces (130 wide in average). The root-means-square (rms) roughness is below 0.2 nm in all samples. Thinner films have high density of two-dimensional islands near the upper steps (see zooms in the insets of Figures 2a and 2b). Two-dimensional islands are not observed in thicker films, although the surface of these films remains very flat and the line topographic profiles show height variation of less than 1 nm over a distance of 5 μm (Supporting Information S4). Terraces are not distinguished in films on Si(001) (Supporting Information S5), but they are very flat, with rms around 0.2 - 0.3 nm, and height profiles within a range of 2 nm along a distance of 5 μm.

Figure 3a shows polarization loops of the films on STO(001). The loops were measured by the DLCC method (see Experimental Section), and thus they include ferroelectric and dielectric contributions, the latter causing the slope evident at high fields. All films are ferroelectric, with remanent polarization $P_r$ in the 20-30 μC/cm$^2$ range. Due to the low thickness of the films and the large electric field applied, there is an influence of leakage, particularly for the t = 4.8 nm film and positive voltage. The effect of leakage is twofold: it can add to switching current and polarization can be overestimated, but on the contrary, it does not allow full saturation of the loop and therefore the polarization is underestimated. Polarization loops measured at increasing voltage (Supporting Information S6) confirm that leakage is less prominent while increasing thickness and the loops are saturated for high-applied voltage. In addition, residual leakage subtraction performed using reported equations[32] shows that its contribution is 2 μC/cm$^2$, which it is taken as a sensitivity limit (Supporting Information S7). Remanent polarization is represented in Figure 3b (blue solid up triangles) as a function of thickness. $P_r$ is ~20 μC/cm$^2$ in the thinner film, t = 4.8 nm, increases to ~28 μC/cm$^2$ in the t = 6.2 nm film, and for thicker films reduces with thickness up to ~21 μC/cm$^2$ in the t = 13 nm film. Figure 3b includes $P_r$ values (green squares) of equivalent epitaxial undoped HZO films on STO(001)[15] Both series show the same dependence on thickness, with a maximum polarization for films thinner than 10 nm, and having La-doped films slightly higher $P_r$ than undoped films. Both series show the same dependence on thickness, with a maximum polarization for films thinner than 10 nm, and having La-doped films higher $P_r$ than undoped films.

The polarization loops of the films on Si(001) are presented in Figure 3c. Loops corresponding to thinner films are not saturated due to leakage (see measurements at varying maximum voltage in Supporting Information S8). Remanent polarization $P_r$ decreases monotonically with thickness (solid red down triangles in Figure 3b) from >30 μC/cm$^2$ to 21 μC/cm$^2$. Similar high $P_r$ in t< 10 nm films was reported for undoped HZO films integrated epitaxially on Si(001) using yttria-stabilized zirconia (YSZ) buffer layers (pink open circles in Figure 3b).[22] Polarization decreases with thickness for both doped and undoped HZO films, but the reduction is less for La-doped films. Figure 3b also shows for films on STO(001) that the reduction of polarization with thickness in films thicker than around 10 nm is less for La-doped films than for undoped ones. The higher remanent polarization in thicker films could be due to the lower amount of monoclinic phase in La-doped HZO films. Undoped HZO films with very large remanent polarization were also integrated epitaxially on STO buffered Si(001) wafers, but the dependence on thickness was not reported[16] The $P_r$ value reported for a t ~ 8 nm HZO film is included in Figure 3b (orange diamond).

The $E_c$ of the La:HZO films (Figures 3a and 3c) is very large. $E_c$ is represented as a function of thickness in Figure 3d. As thickness increases from 4.8 to 13 nm, $E_c$ decreases from 3.7 to 2.3

MV/cm in films on STO(001) (blue up triangles), and from 4.2 to 2.5 MV/cm in films on Si(001) (red down triangles). $E_c$ of films on both substrates depends on thickness following the $E_c \propto t^{-2/3}$ scaling law. This phenomenological law is common in ferroelectric perovskites,[33] but is not observed in polycrystalline ferroelectrics based on $HfO_2$, which generally show little thickness dependence.[34] The $E_c \propto t^{-2/3}$ scaling has been also reported for epitaxial undoped HZO films.[15, 22] We note that coercive field of both La-doped and undoped epitaxial HZO films is higher than that of equivalent polycrystalline films, being the difference higher for thinner films as polycrystalline films show little thickness dependence. We include in Figure 3d reported data for epitaxial HZO films on STO(001) (green squares) and YSZ buffered Si(001) (pink circles). $E_c$ is seen to be greater in doped HZO films than in undoped HZO films. This is in contrast to the reduction of $E_c$ by approximately 30% in polycrystalline HZO films doped with same 1% mol La content.[5-6] Although the cause of $E_c$ reduction in polycrystalline La doped films is unknown, the possible presence of tetragonal phase was suggested,[6] but this phase apparently is not present in the epitaxial films on LSMO(001) electrodes.

Leakage current curves for La-doped HZO films on STO(001) are shown in Figures 4a. Leakage is low, even in the sub-5 nm film, which is around $10^{-6}$ A/cm$^2$ at 1 MV/cm and $10^{-5}$ A/cm$^2$ at 2 MV/cm. Leakage decreases with film thickness, and the values in the t = 13 nm film are about one order of magnitude less than in the t = 4.8 nm film. Films on Si(001) are slightly more leaky, with the leakage current of the thinner film around $3\times10^{-6}$ A/cm$^2$ at 1 MV/cm and $5\times10^{-5}$ A/cm$^2$ at 2 MV/cm. Similarly to the films on STO(001), leakage is significantly reduced with thickness. Figure 4c shows the thickness dependence of the current leakage at 2 MV/cm of films on STO(001) (blue up triangles) and Si(001) (red down triangles). The graph also includes leakage data from equivalent epitaxial undoped HZO films on STO(001) (green squares).[15] It is observed that the leakage current of epitaxial La-doped HZO films is considerably reduced, 1 - 2 orders of magnitude, compared to equivalent undoped HZO films.

Figure 5 shows the current-voltage (I-V) curves and the polarization loops of the films on STO(001), in pristine state and measured after 10 and 100 cycles. The t = 4.8 nm film shows in pristine state a double peak in the I-V curve (Figure 5a) and a pinched polarization loop (Figure 5b). The two peaks get closer after 10 cycles, and after 100 cycles, they merge in a single narrower peak, and the corresponding polarization loop is not pinched. In the t = 6.3 nm film (Figure 5c), the two peaks are closer in the pristine state, and after 10 pulses, only a small secondary peak on the positive axis is distinguished at a higher voltage than the main peak. With more cycles, the amplitude of the main peak has not changed, while the second peak has disappeared. The corresponding ferroelectric loops (Figure 5d) show a significant increase of polarization with respect to the pristine state. In contrast, no differences in the I-V curves were observed between the pristine state and the 10 times cycled t = 8.3 nm film (Figure 5e). There is a single peak that features a shoulder on the higher voltage side. The switching peak has a similar amplitude after 100 cycles, but without shoulder. Consistent with this, the polarization loops (Figure 5f) show no significant differences and there is only a very slight reduction in polarization after 100 cycles. Finally, in the thickest film, t = 13 nm, there is a single peak in the pristine state, which decreases in amplitude by cycling (Figure 5g). The resulting polarization loops (Figure 5h) show a reduction in remanent polarization with cycling. Therefore, the behavior of epitaxial La:HZO films with cycling is highly dependent on thickness. The two thinnest epitaxial films on Si(001) show a similar wake-up effect (Figure 6). The wake-up effect, which occurs recurrently in polycrystalline hafnia films, does not occur in epitaxial La:HZO films thicker than about 8 nm. In contrast, the films t = 6.3 and 4.8 nm waken after around 10 and 100 cycles, respectively. The wake-up effect is more evident than in same thickness epitaxial undoped HZO films, where the effect is less evident. In La-doped polycrystalline HZO films, wake-up effect increases strongly compared to undoped films.[5-7] Indeed, wake-up effect in polycrystalline La-doped HZO films can extent to a larger number of about $10^7$ cycles,[6] which is one of the main drawbacks of La-doped HZO films. Therefore, the strong reduction or suppression of wake-up effect in epitaxial La:HZO films with respect to polycrystalline La:HZO films is relevant.

Next, we investigate the endurance of the films. The capacitors were cycled, up to a maximum number of $5 \times 10^{10}$ cycles, or until breakdown (denoted by empty symbols) or a fatigued state with remanent polarization reduced to below 1.5 µC/cm$^2$. A set of endurance measurements was done for each sample by cycling the capacitors with pulses of different amplitude. Figure 7a presents the polarization loops of the t = 4.8 nm film on STO(001), measured at 6.5 MV/cm. The initial polarization increases after 10 cycles due to the wake-up effect discussed above, and additional cycling progressively reduces the polarization. $2P_r$ is plotted against the number of cycles in Figure 7b (red circles). The maximum $2P_r$ = 16.6 µC/cm$^2$ after 10 cycles decreases to 7.5 µC/cm$^2$ after $10^9$ cycles, before the capacitor breaks. Therefore, endurance of the capacitor is limited first by fatigue and finally by breakdown. Figure 7b includes endurance measurements under other electric fields. With high field of 7.6 MV/cm (blue triangles), the initial remanent polarization is greater, $2P_r$ = 19 µC/cm$^2$, and increases to a maximum value of 22.5 µC/cm$^2$ after 10 cycles. With additional cycles, the capacitor is fatigued, but retains $2P_r$ = 13.8 µC/cm$^2$ after $10^7$ cycles before hard breakdown. For the low cycling field (5.4 MV/cm, black squares), $2P_r$ is low in the pristine state, 8.6 µC/cm$^2$, and wake-up effect persists for around $10^2$ cycles. Then $2P_r$ progressively decreases from 11.1 to 3 µC/cm$^2$ after $5 \times 10^{10}$ cycles, without breakdown. Therefore, the number of cycles without breakdown is strongly extended using a low switching voltage. However, polarization is reduced and the wake-up effect is more persistent under these conditions, while there is not wake-up for enough high poling voltage. This is in agreement with the improved oxygen diffusion and the reduction of wake-up effect in polycrystalline HfO$_2$-based films increasing amplitude[35] or applied field duration.[36-37] The polarization loops (Figure 7d) of the t = 6.3 nm film, measured at 5.1 MV/cm, show wake-up effect during some cycles, and the maximum $2P_r$ = 24.8 µC/cm$^2$ after 10 cycles is reduced with more cycles to $2P_r$ = 2 µC/cm$^2$ after $10^{10}$ cycles. Endurance is similar for switching field of 4.3 MV/cm (Figure 7e). By increasing the field to 5.8 MV/cm, the initial polarization increases, but hard breakdown occurs after $10^7$ cycles. Thicker films do not show wake-up, and polarization decreases continuously with cycling. Hard breakdown takes place in the t = 8.3 nm film (Figures 7g and 7h) after $2 \times 10^7$ cycles at 5.4 MV/cm or $2 \times 10^9$ cycles at 4.9 MV/cm. Breakdown did not occur using a lower switching field of 4.3 or 4.0 MV/cm, but $2P_r$ decreased to 3.4 µC/cm$^2$ after $10^{10}$ cycles. Finally, the t = 13 nm film (Figures 7j and 7k) shows breakdown after $10^9$ cycles at 4.0 or 3.6 MV/cm, while for a lower applied field of 3.2 MV/cm the polarization decreased to $2P_r$ = 2.3 µC/cm$^2$ after $10^7$ cycles. Similar strong robustness against breakdown when the switching field is reduced was reported for polycrystalline doped HfO$_2$ films.[38] Breakdown in epitaxial films, however, occurs at much higher fields than in polycrystalline films. The endurance measurements of the La:HZO films on Si(001) are summarized in Figure 8. The influence of the switching voltage on endurance follows the observed dependences of the films on STO(001), although the maximum endurance of films on Si is limited to around $10^9$ cycles.

Figure 7 shows that La:HZO films on STO(001) suffer fatigue at all switching voltages, and similar behavior is observed in films on Si(001) (Figure 8). We have investigated the influence of the cycling voltage on fatigue. The normalized endurance (Supporting Information S9) shows the same percentage decrease, that is, fatigue of each film does not depend on the value of the electric field used to test it. Similar independence of the electric field had been observed in perovskite ferroelectrics,[39-40] and it was proposed that there is an influence of electric field amplitude only for very short cycling pulses.[41] In the case of doped $HfO_2$ films, both clear dependence on the amplitude of the electric field[42] and no obvious dependence (as published graphs suggest, since normalized polarization is not usually plotted)[35, 43] have been reported. The independence of fatigue on switching electric field in the epitaxial La:HZO films allows direct comparison of fatigue in films of different thickness. The normalized polarization of the La:HZO films on STO(001) (Figure 9a) and Si(001) (Figure 9b) shows little thickness dependence on fatigue. In contrast, a monotonic increase of fatigue with thickness was reported of undoped epitaxial films.[22] Therefore, doping with La mitigates the degradation of endurance of ferroelectric hafnia with increasing thickness. Increasing thickness, the minority monoclinic phase increases for both La-doped and undoped HZO epitaxial films, but the amount is smaller in the La-doped films. This suggests that the reduction of paraelectric phase could be a relevant factor to mitigate fatigue (Figure 9b) and improve endurance (Figure 9d).

The endurance study summarized in Figure 7 includes the current leakage as a function of the number of cycles. The evolution of leakage during the cycling of the t = 4.8, 6.3, 8.3 and 13 nm films is presented in Figures 7c, 7f, 7i and 7l, respectively. Leakage remains constant for a certain number of cycles, but increases sharply by several orders of magnitude after a threshold. The defects generated by cycling probably accumulate at the boundaries between the grains and the crystal variants. The observed threshold in the number of cycles suggests that the new defects only have an impact on the leakage when their density is high enough to allow percolation. The threshold occurs in each sample at a lower number of cycles as the switching voltage is higher (Figure 9c). Moreover, the leakage current increases after fewer cycles with a similar electric field as the film becomes thicker. On the other hand, the dependence of $2P_r$ on the number of cycles in Figure 7 shows that hard breakdown (marked with empty symbols) occurs after fewer cycles and a smaller switching electric field as the films becomes thicker. This is evidenced in Figure 9d, where the endurance of the four films is plotted as a function of the switching field. A similar thickness dependence had been observed in polycrystalline HZO films,[44] although breakdown fields in epitaxial films are much higher. In epitaxial La:HZO films, endurance was limited by hard breakdown (open symbols), in most tested capacitors. However, in capacitors switched with a low electric field, the low polarization of fatigued capacitors (solid symbols) limits endurance.

Polarization loops generally show different $E_c$ on the negative and positive axes. The imprint field ($E_{imp}$) of the films on STO(001) is represented as a function of the number of cycles in Figure 10. The dependences indicate that $E_{imp}$ is not only caused by the different work function of the Pt and LSMO electrode. An inhomogeneous distribution of oxygen vacancies can create an internal field that would increase the Schottky barrier height at one interface and decrease it at the other. The loops of the two thickest films (t = 8.3 nm and 13 nm) are displaced towards the negative axis of the electric field, with the $E_{imp}$ directed from the upper Pt towards the lower LSMO electrode. The $E_{imp}$, around 300 kV/cm, decreases with the number of cycles, up to 130 kV/cm in the t = 8.3 nm film (blue up triangles) and up to negligible imprint in the t = 13 nm film (green down triangles). The behavior clearly differs in the two thinnest films, in which the $E_{imp}$ in the pristine state is directed from the lower LSMO towards the upper Pt electrode, its magnitude being close to 400 and 200 kV/cm in t = 4.8 and 6.3 nm films, respectively. After 10 cycles, the $E_{imp}$ reverses its direction. In the t = 6.3 nm film (red circles) $E_i \sim$ 300 kV/cm decreases to less than 100 kV/cm after $10^{10}$ cycles, while in the t = 4.8 nm film (black squares) the smallest $E_i \sim$ 80 kV/cm quickly decreases to almost vanish during additional cycling. There is a correlation between the large $E_{imp}$ in the two thinnest films in their pristine state that decreased rapidly after only about 10 cycles and the wake-up effect of these samples. In general, wake-up is considered to be caused by oxygen vacancies diffusing under the action of the electric field, allowing domain depinning and increased polarization.[4, 35-36] $E_{imp}$ in the pristine state decreases as the vacancies diffuse under the cumulative action of the electric field cycles. Reduction of internal field during wake-up was also observed in polycrystalline hafnia.[45-46] In the case of the epitaxial La-doped films on LSMO, pinned dipoles at the interface could be another contribution to $E_{imp}$ in the pristine state, as has been reported for ferroelectric perovskites on LSMO.[47] These dipoles can pin ferroelectric domains, reducing the overall polarization. The large positive imprint field in the two thinnest La-doped HZO films can be attributed to these interface dipoles, since its magnitude is larger as thinner the ferroelectric layer is. The observed high reduction of imprint field with few cycles in these two thinnest films can be due to charge injection, causing a space charge layer.[48-49] It would cancel the field created by the interfacial dipoles and suppress the pinning of ferroelectric domains, thus waken-up the capacitors.

Figures 11a-d show the polarization retention of the films on STO(001), measured in pristine state at room temperature. Retention of the t = 4.8 nm film is highly dependent on the direction of poling, with the remanent polarization extrapolated to 10 years 2 µC/cm$^2$ for positive voltage and > 10 µC/cm$^2$ for negative poling. The asymmetry is less in thicker films, and the extrapolated remanent polarization for both positive and negative poling is high in all films. Data collected at 85 °C in representative t = 8.3 nm samples on STO and Si are plotted in Figures 11e,f. Data show that the retention is still high at high temperature with extrapolated polarization at 10 years larger than 5 µC/cm$^2$. We have quantified retention by fitting the measured data to the $P_r=P_0 \, t_d^{-k}$ equation (dashed lines in Figures 11a-d), where $t_d$ refers to the time after poling.[50-51] The graph of the exponent k against thickness in Figure 9g confirms that the asymmetry is much greater in the t = 4.8 nm film than in the other films including data measured at 85 °C. The corresponding graph for the films on Si(001) is shown in Figure 11h (retention data is in Supporting Information S10). The results confirm the observations on STO: strong asymmetry in the film of thickness less than 5 nm and very long retention in the thicker films for both poling directions. Imprint field, current leakage, and depolarizing field are expected to determine the influence of poling direction and film thickness on retention. Imprint field, critical in the asymmetry, may depend on the presence of monoclinic phase, more important in thicker films, and also on the characteristics of the voltage pulses applied (amplitude, time and polarity).[7] Current leakage decreases monotonically with thickness, which favors a better retention in thicker films. However, the effect of leakage depends on spatial inhomogeneities, mainly due to grain boundaries, and on the possible coexistence of electronic and ionic contributions with different time scales. Finally, thinner films should be more unstable due to the larger depolarization field, although their larger $E_c$ will have the opposite effect. Nevertheless, the excellent properties of the epitaxial La doped HZO films are achieved with endurance that exceeds $10^{10}$ cycles, and the films simultaneously exhibit good retention for more than 10 years under the same poling voltage (Supporting Information S11). The poling voltage of around 3 V can allow the use of the capacitors in a ferroelectric random access memory.[52]

CONCLUSIONS

In conclusion, orthorhombic La-doped HZO films have been epitaxially grown on STO(001) and Si(001) substrates, and their main ferroelectric properties (polarization, endurance and retention) have been determined. Doping does not reduce the $E_c$ of HZO epitaxial films, but the leakage current is substantially reduced. Ferroelectric polarization, particularly in films over 10 nm thick, is greater than that of undoped films. The wake-up effect, a serious drawback in polycrystalline La-doped films, is limited to approximately 100 cycles in epitaxial films and only occurs in thinner films. The films exhibit fatigue, but endurance exceeds $10^{10}$ cycles, and the films simultaneously exhibit very high retention for more than 10 years. This demonstrates that there is not an intrinsic dilemma between endurance and retention in La-doped HZO films.

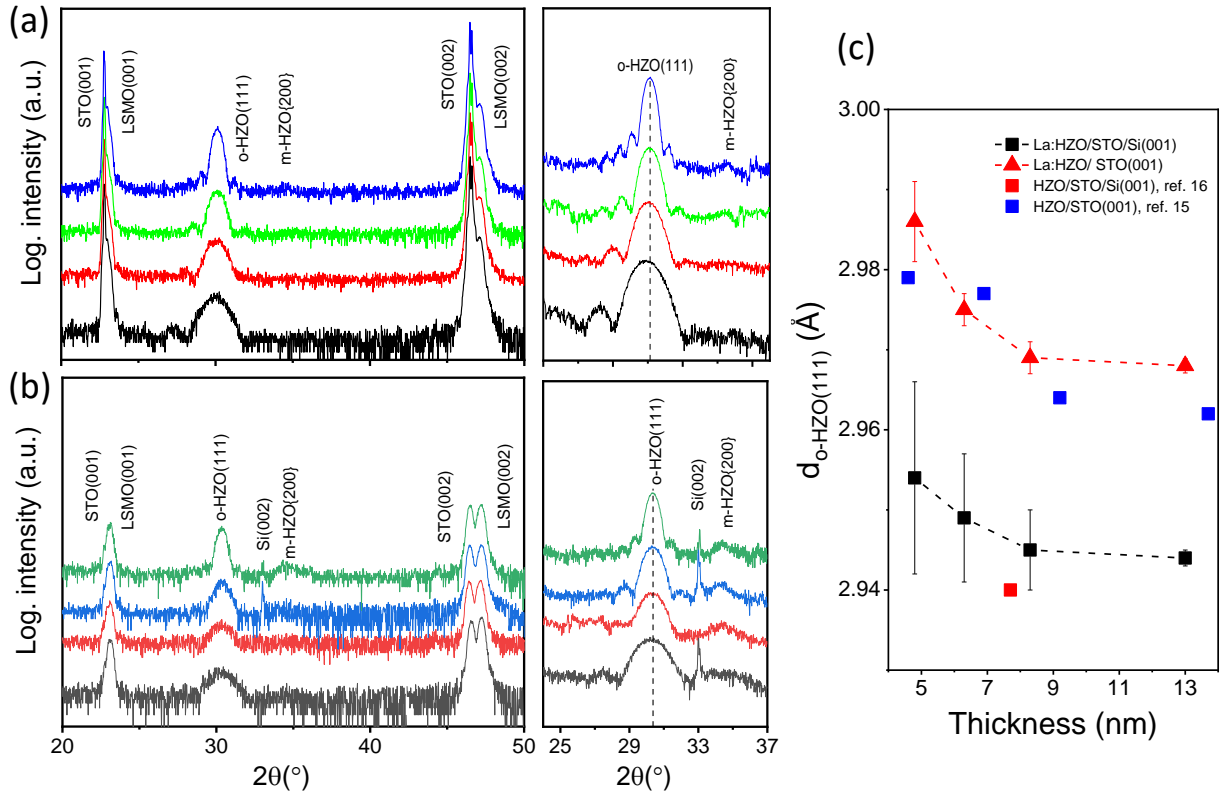

**Figure 1.** XRD θ-2θ scans of La:HZO films on (a) LSMO/STO(001) and (b) LSMO/STO/Si(001). Right panels: scans acquired with longer time. (c) Out-of-plane o-HZO(111) lattice distance of La:HZO films on LSMO/STO(001) (red triangles) and LSMO/STO/Si(001) (black squares), plotted as a function of thickness. Out-of-plane lattice distance values of epitaxial undoped HZO on LSMO/STO(001) (blue squares)[15] and LSMO/STO/Si(001) (red square)[16] are plotted for comparison.

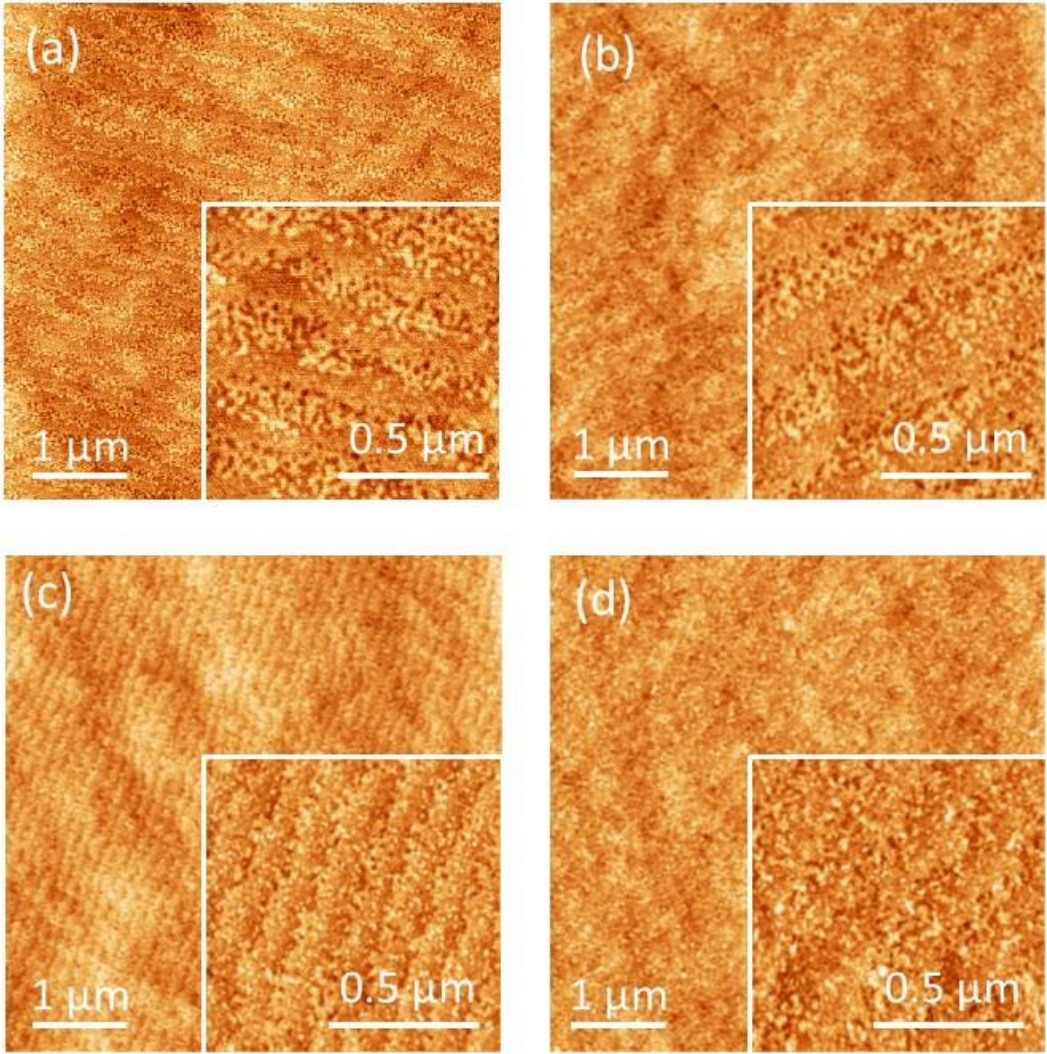

**Figure 2.** 5 μm x 5 μm AFM topographic images of (a) t = 4.8 nm, (b) t = 6.3 nm, (c) t = 8.3 nm, and (d) t =13 nm films on STO(001). Insets: topographic images of 1 μm x 1 μm scanned areas.

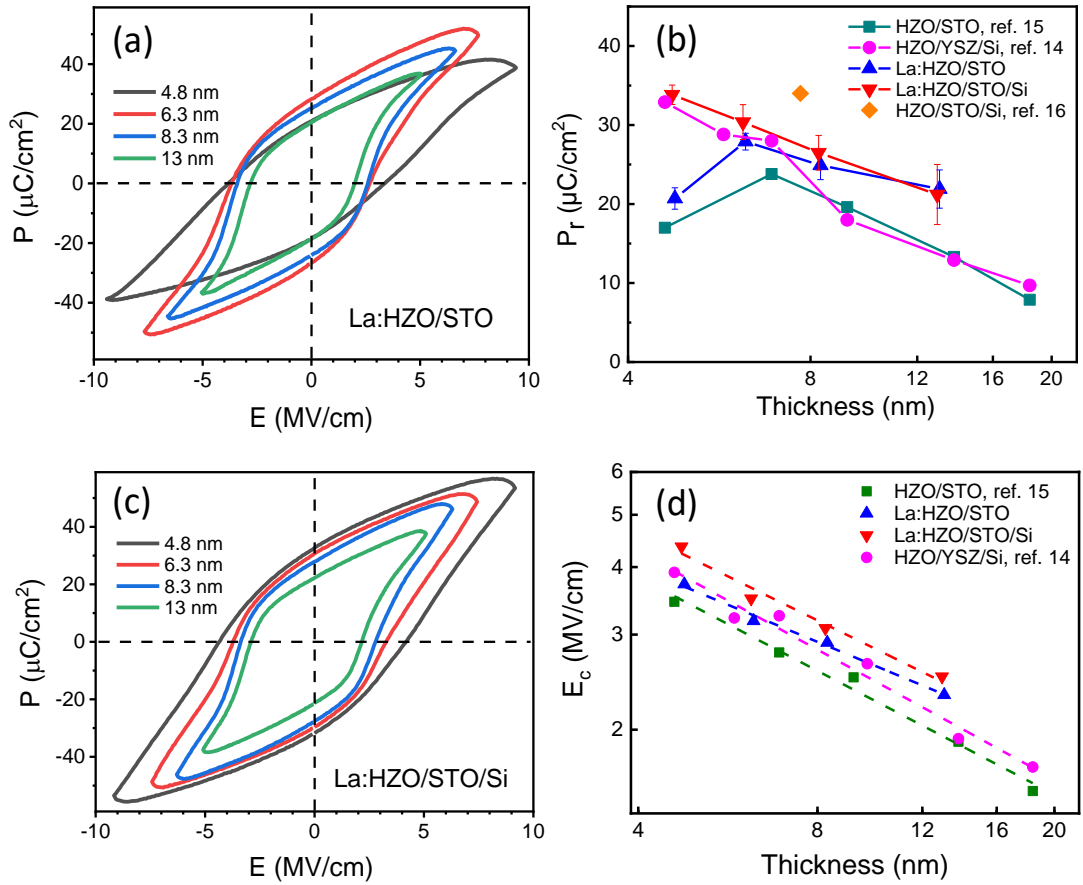

**Figure 3.** Polarization loops of La:HZO films on STO(001) (a) and Si(001) (c). Dependence of remanent polarization (b) and coercive field (d) on thickness for La:HZO films on STO(001) (blue up triangles) and Si(001) (red down triangles). Data reported for epitaxial undoped HZO films on STO(001) (green squares) and YSZ-buffered Si(001) (pink circles) are included. Error bar on remanent polarization corresponds to the standard deviation among around 10 different measured capacitors.

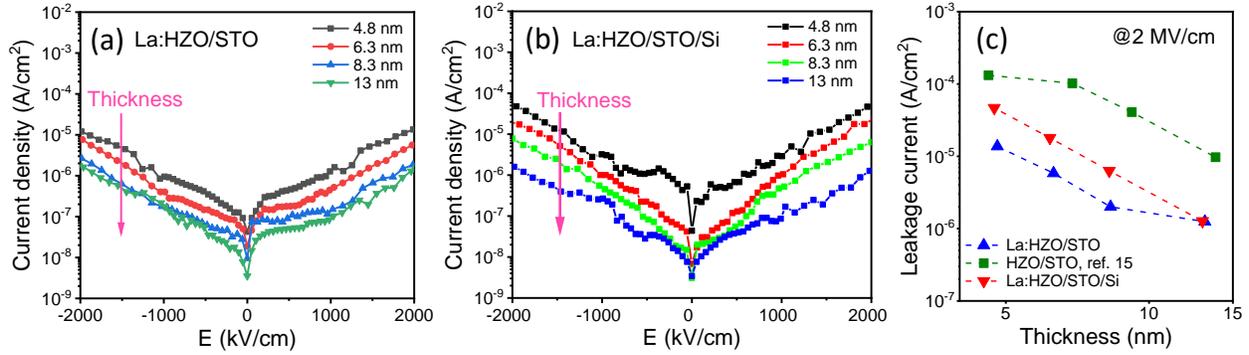

**Figure 4.** Leakage current of La:HZO films on (a) STO(001) and (b) Si(001). (c) Dependence of leakage current on thickness for La:HZO films on STO(001) (blue up triangles) and Si(001) (red down triangles). Reported data for epitaxial undoped HZO films on STO(001) (green squares) is included.

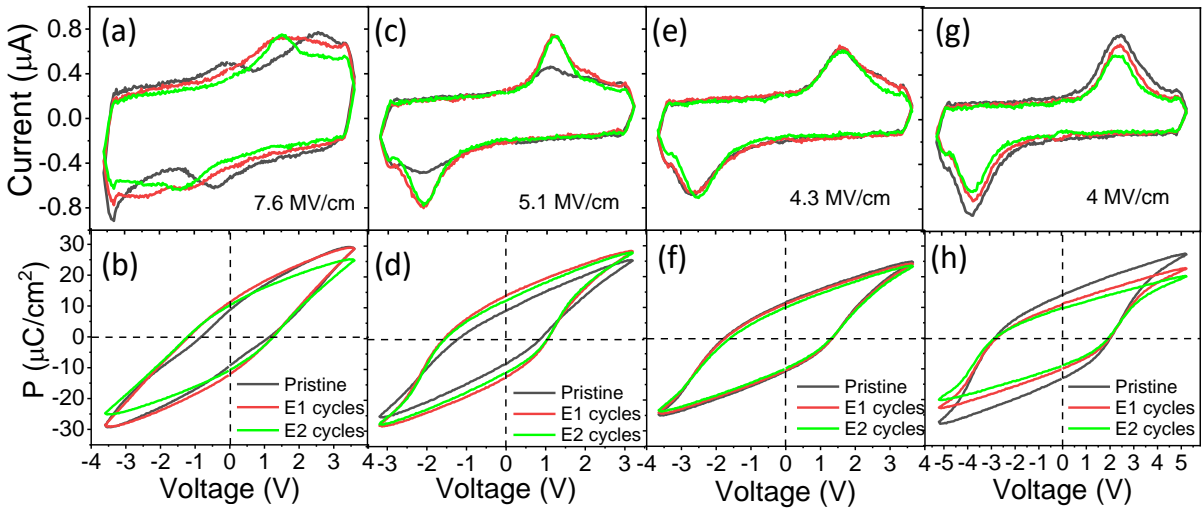

**Figure 5.** (a) Current - voltage (I-V) curves measured in pristine state and after 10 and 100 cycles and (b) corresponding polarization - voltage (P-V) loops for the t = 4.8 nm film on STO(001). I-V curves and P-V loops for the t = 6.3, 8.3 and 13 nm films are shown in panels (c-d), (e-f) and (g-h), respectively.

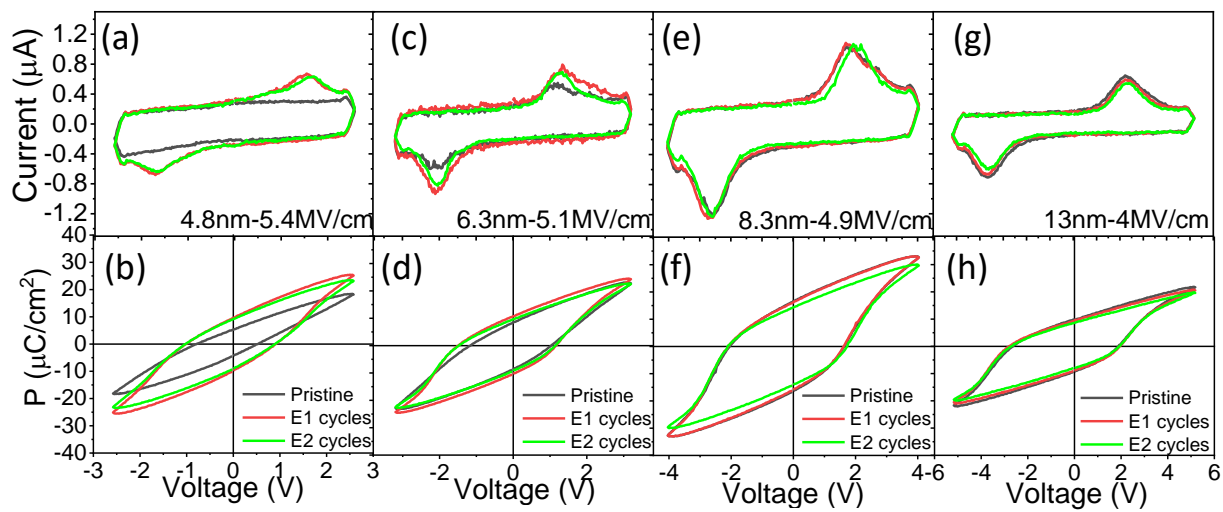

**Figure 6**. (a) Current - voltage (I-V) curves measured in pristine state and after 10 and 100 cycles and (b) corresponding polarization - voltage (P-V) loops for the t = 4.8 nm film on Si(001). I-V curves and P-V loops for the t = 6.3, 8.3 and 13 nm films are shown in panels (c-d), (e-f) and (g-h), respectively.

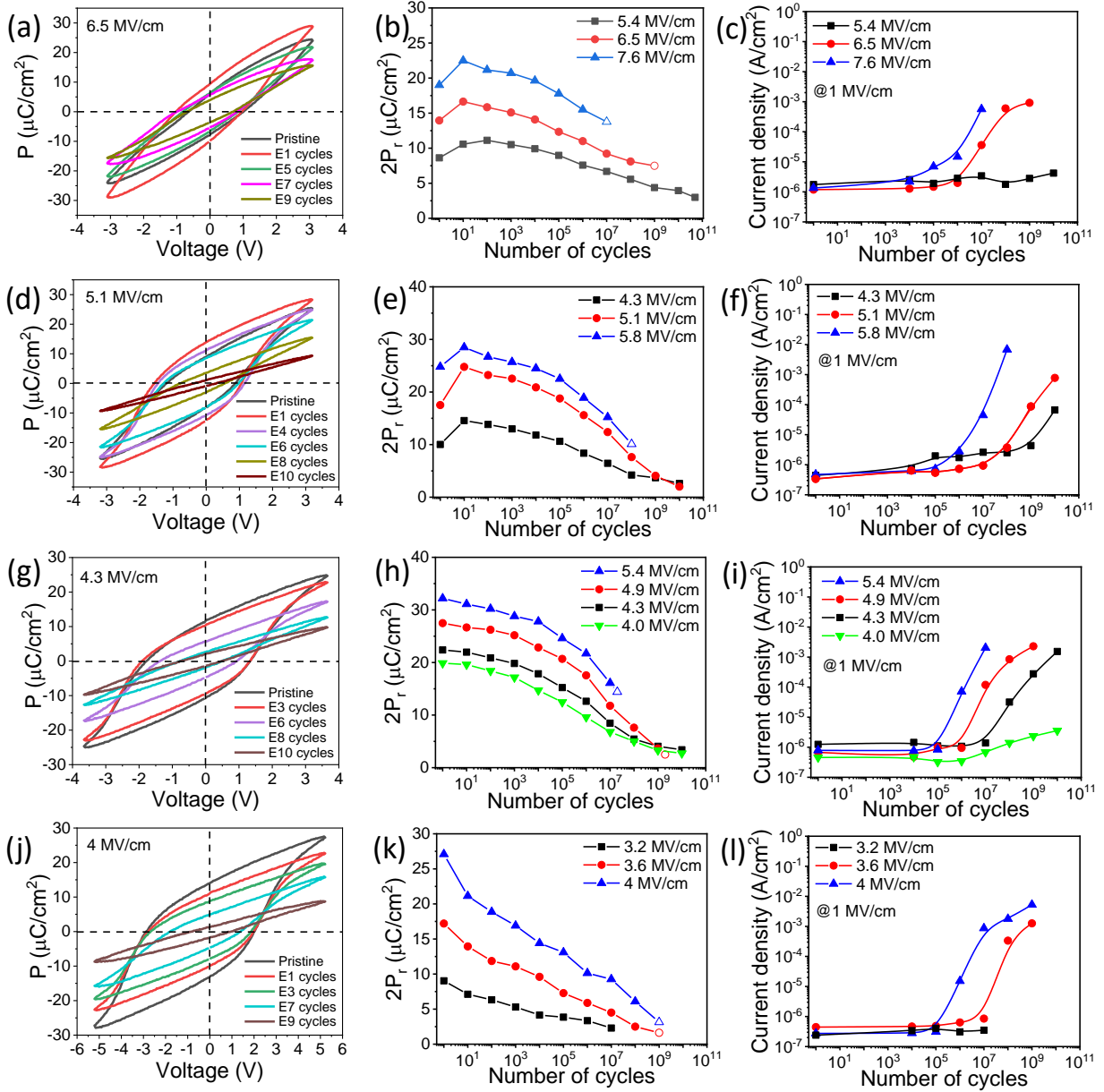

**Figure 7.** (a) Polarization - voltage P-V loops, (b) endurance and (c) evolution of current leakage with number of cycles for the t = 4.8 nm film on STO(001). The P-V loops, endurance and evolution of current leakage with number of cycles corresponding to the t = 6.3 nm, 8.3 nm and t = 13 nm films are shown in panels (d-f), (g-i) and (j-l), respectively. Empty symbols indicates last measured data point before breakdown.

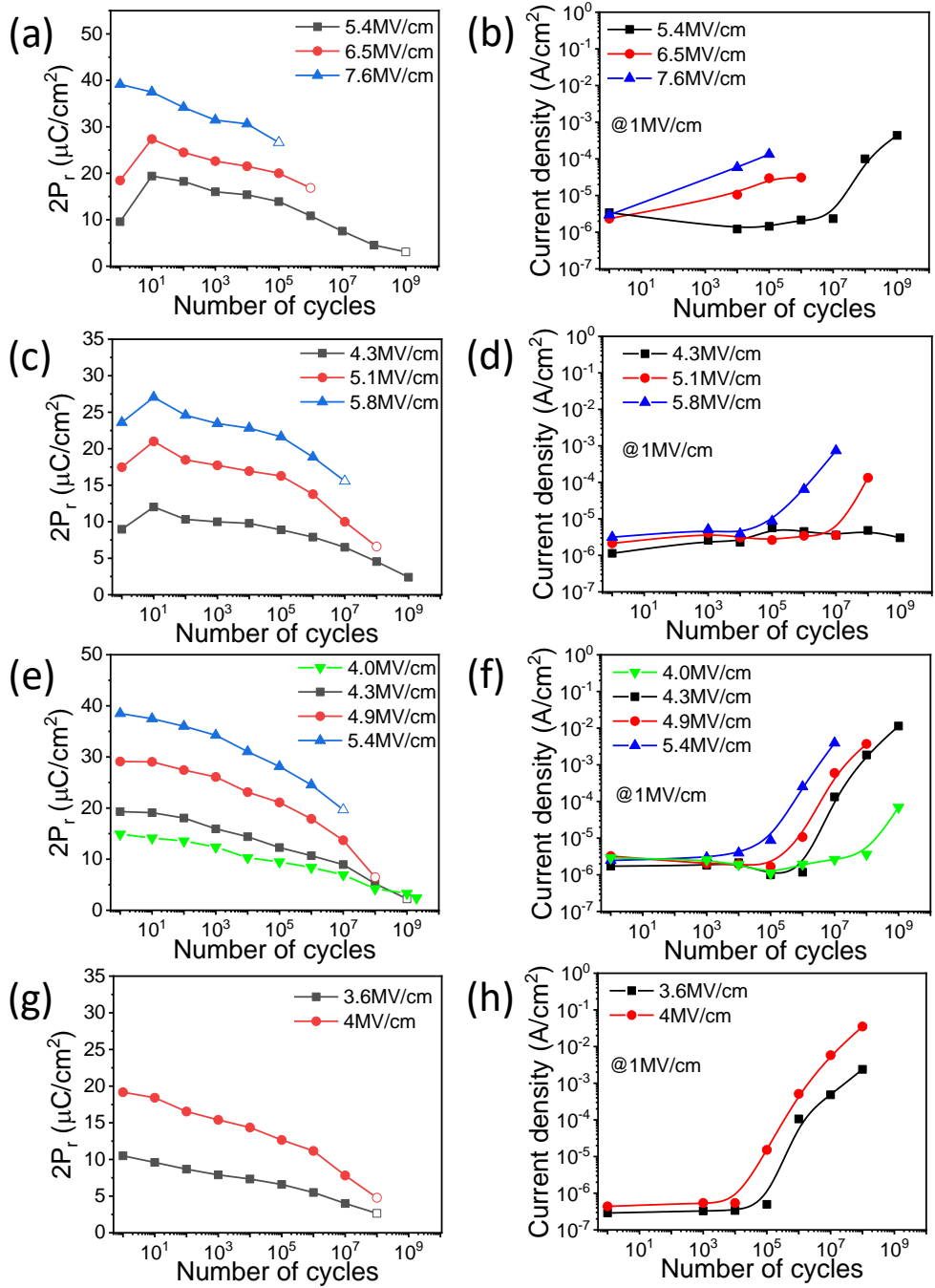

**Figure 8**. (a) Endurance and (b) variation of current leakage measured at 1MV/cm with number of cycles for the t = 4.8 nm film on Si(001). The endurance and variation of current leakage with number of cycles corresponding to the t = 6.3 nm, 8.3 nm and t = 13 nm films are shown in panels (c-d), (e-f) and (g-h), respectively.

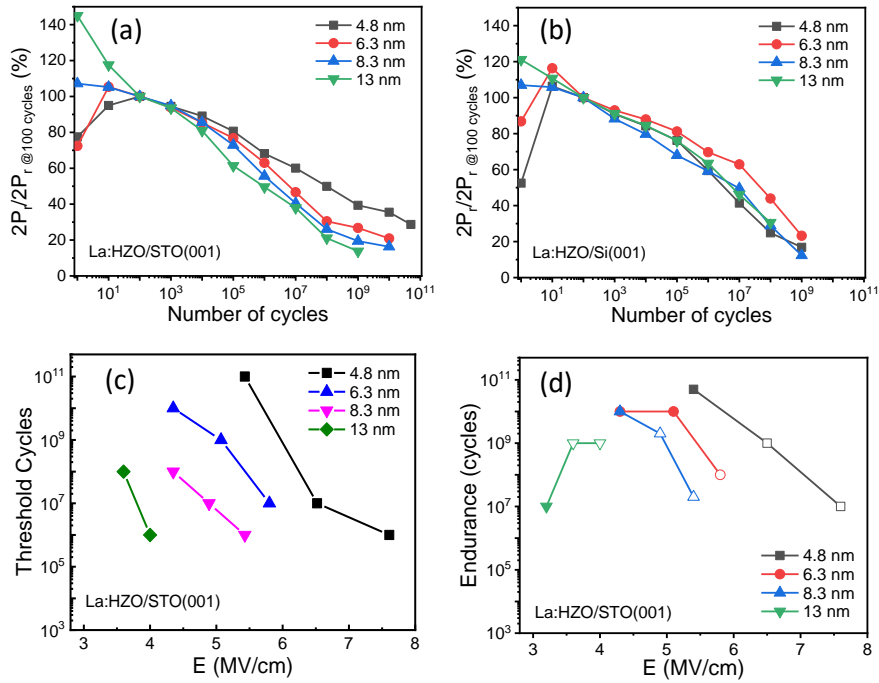

**Figure 9.** Endurance data normalized to $2P_r$ at 100 cycles of films on (a) STO(001) and (b) Si(001). (c) Map of the threshold of number of cycles for abrupt leakage increase as a function of electric field and thickness of films on STO(001). (d) Map of endurance as a function of electric field and thickness of films on STO(001). Empty symbols denote breakdown.

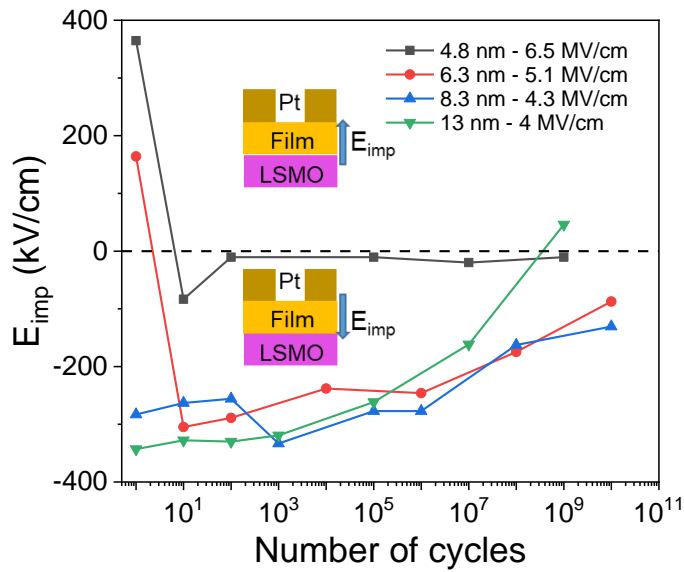

**Figure 10.** Variation of imprint field ($E_{imp}$) with the number of cycles for the films on STO(001).

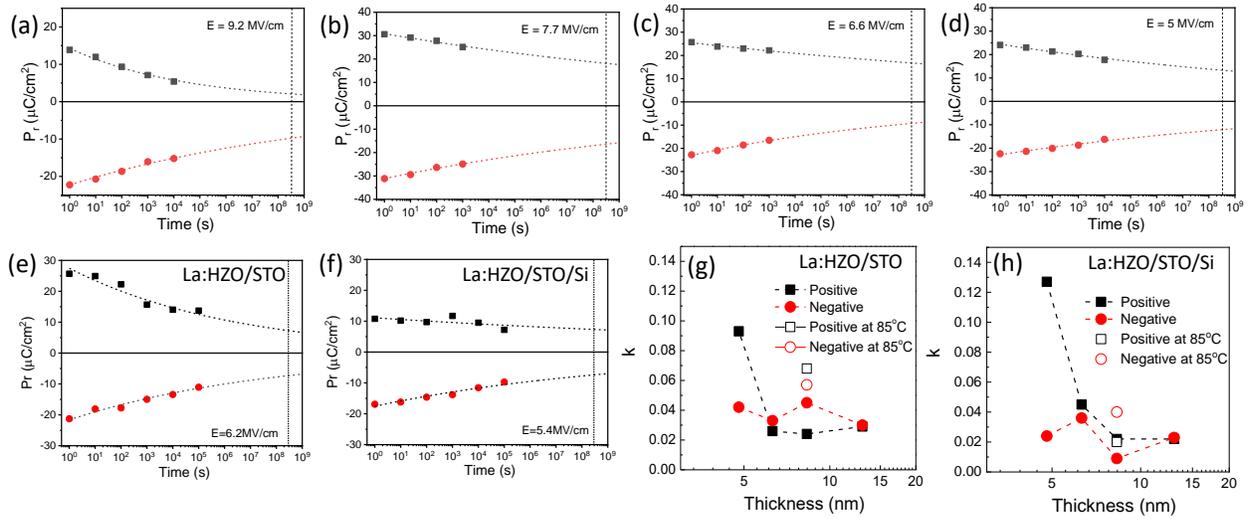

**Figure 11.** Polarization retention at room temperature of (a) t = 4.8 m, (b) t = 6.3 nm, (c) t = 8.3 nm and (d) t = 13 nm films on STO(001). Polarization retention at 85 °C of films with t = 8.3 nm on (e) STO(001) and (f) Si(001). Lines are fits to $P_r = P_0 t_d^{-k}$ equation for positive and negative poling. The vertical dashed lines mark a time of 10 years. k parameter plotted for positive and negative poling as a function of thickness for films on (g) STO(001) and (h) Si(001).

**Supporting Information.**

Simulation of Laue oscillations. XRD pole figures and 2θ-χ frames. Topographic AFM images and height profiles. Polarization loops measured at increasing voltage of films on STO(001) and Si(001). Determination of residual leakage contribution to the polarization loops. Influence of the cycling voltage on fatigue of films on STO(001) and Si(001). Polarization retention of the films on Si(001). Polarization retention of films on STO(001) poled at varying field.

**Corresponding Authors**

* Ignasi Fina: ifina@icmab.es and Florencio Sánchez: fsanchez@icmab.es

**Funding Sources**

From the Spanish Ministerio de Ciencia e Innovación, through the "Severo Ochoa" Programme for Centres of Excellence in R&D (SEV-2015-0496) and the MAT2017-85232-R (AEI/FEDER, EU), PID2019-107727RB-I00 (AEI/FEDER, EU), and MAT2015-73839-JIN projects, and Ramón y Cajal contract RYC-2017-22531. From Generalitat de Catalunya (2017 SGR 1377). China Scholarship Council (CSC), grant No. 201807000104.

ACKNOWLEDGMENT

Financial support from the Spanish Ministerio de Ciencia e Innovación, through the "Severo Ochoa" Programme for Centres of Excellence in R&D (SEV-2015-0496) and the MAT2017-85232-R (AEI/FEDER, EU), PID2019-107727RB-I00 (AEI/FEDER, EU), and MAT2015-73839-JIN projects, and from Generalitat de Catalunya (2017 SGR 1377) is acknowledged. IF acknowledges Ramón y Cajal contract RYC-2017-22531. TS is financially supported by China Scholarship Council (CSC) with No. 201807000104. TS work has been done as a part of his Ph.D. program in Materials Science at Universitat Autònoma de Barcelona.